\documentclass[sigconf]{acmart}
\usepackage{booktabs}
\usepackage{multirow}
\usepackage{array}
\usepackage{graphicx}
\usepackage{xcolor}
\usepackage{placeins}
\usepackage{enumitem}

\AtBeginDocument{%
  }

\setcopyright{none}
\copyrightyear{2026}
\acmYear{2026}

\acmConference[Workshop on Evaluation and
Trustworthiness of Agentic AI @ KDD'26]{Workshop on Evaluation and Trustworthiness of Agentic AI, co-located with the 32nd ACM SIGKDD Conference on Knowledge Discovery and Data Mining}{August 9--13, 2026}{Jeju, Korea}

\renewcommand\footnotetextcopyrightpermission[1]{}
\begin{document}

%% Title.
\title{A Framework for Evaluating Agentic Skills at Scale}

%% Authors --- placeholders, to be replaced once the final author list
%% is confirmed.
\author{Maksim Shaposhnikov}
\email{max@tessl.io}
\affiliation{%
  \institution{Tessl}
  \city{London}
  \country{United Kingdom}
}

\author{Nicolas Fortuin}
\email{nicolas@tessl.io}
\affiliation{%
  \institution{Tessl}
  \city{London}
  \country{United Kingdom}
}

\author{Simon Stipcich}
\email{simons@tessl.io}
\affiliation{%
  \institution{Tessl}
  \city{London}
  \country{United Kingdom}
}

\author{Maria I. Gorinova}
\email{maria@tessl.io}
\affiliation{%
  \institution{Tessl}
  \city{London}
  \country{United Kingdom}
}

\author{Amy Heineike}
\email{amy@tessl.io}
\affiliation{%
  \institution{Tessl}
  \city{London}
  \country{United Kingdom}
}

\author{Rob Willoughby}
\email{robw@tessl.io}
\affiliation{%
  \institution{Tessl}
  \city{London}
  \country{United Kingdom}
}

\renewcommand{\shortauthors}{Shaposhnikov et al.}

%% Abstract.
\begin{abstract}
  Agent skills --- structured, reusable knowledge artifacts that
  augment LLM agent capabilities --- have been rapidly adopted in
  industry, yet their cross-domain impact and use across commercial
  and open-source models remain under-studied, and no reusable
  methodology exists for evaluating an individual skill.
  In this work, we present an evaluation framework that lets a skill author construct
  realistic tasks to rigorously assess the aspects of a skill that
  matter most to them, and that estimates skill utility by solving
  those tasks. Further, we apply our evaluation approach at scale to
  500 real-world skills, generating 1{,}000 tasks derived from the
  skills' content, along with instruction-following and
  goal-completion scoring rubrics. Using these metrics, we evaluate how
  19 agent--model configurations, both proprietary and open-source,
  perform on the tasks. Our results show that models vary widely in
  how closely they adhere to the instructions encoded in skills,
  leading to substantial differences in their performance gains.
  Furthermore, we show that access to a skill significantly changes
  model behavior compared to the no-skill setup, providing an essential
  mechanism for encoding opinionated workflows into LLM agents. We release our evaluation dataset to support future work on
  agent skills. \footnote{\url{https://huggingface.co/datasets/tesslio/task-evals-for-skills}}
\end{abstract}

%% Teaser figure.
\begin{teaserfigure}
  \centering
  \includegraphics[width=\textwidth]{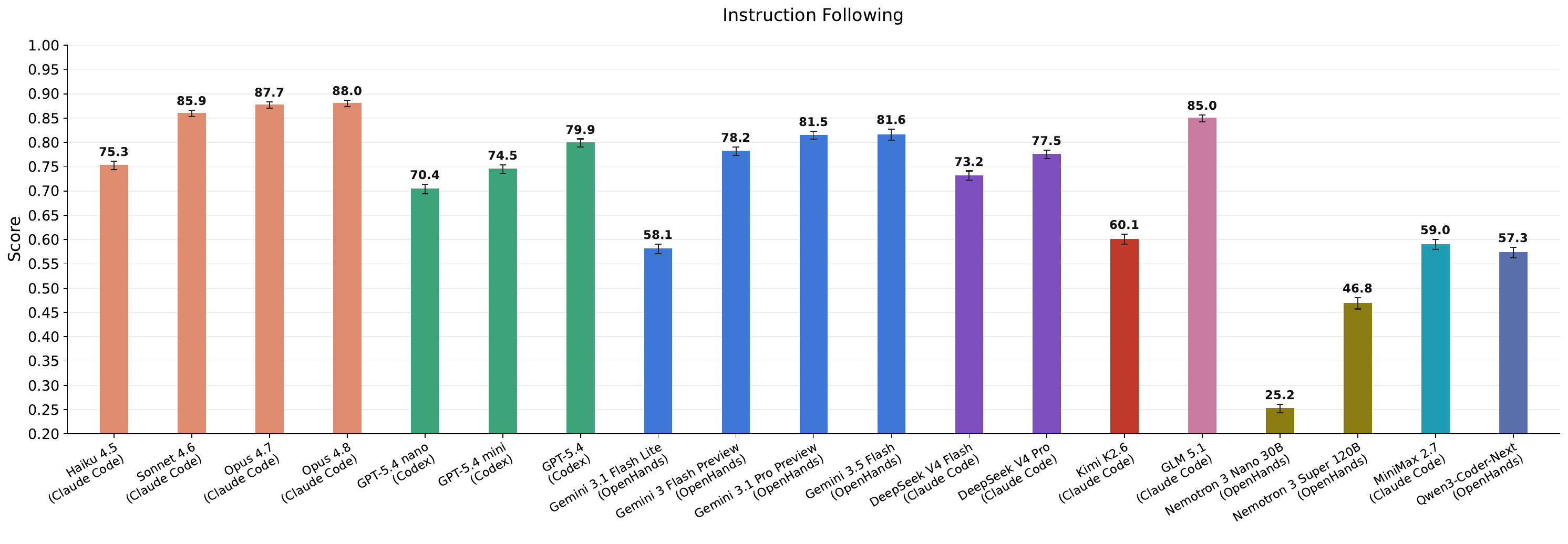}
  \caption{Instruction-following score across every evaluated
    agent--model configuration on our evaluation benchmark of coding tasks that require access to the skill.
    Models vary substantially in how closely they adhere to the instructions encoded in a skill. The
    most expensive proprietary frontier models --- Opus 4.8 (88.0) and
    Opus 4.7 (87.7) --- achieve the highest scores, while the
    open-weights GLM 5.1 reaches a comparable 85.0 at a fraction of
    the cost. By contrast, Kimi K2.6, MiniMax 2.7, Qwen3-Coder-Next,
    and Gemini 3.1 Flash Lite all cluster around 57--60 --- roughly
    25--30 points below the frontier --- and the Nemotron family lags
    by an even wider margin.}
  \Description{Bar chart titled ``Instruction Following'' showing the
    with-skill instruction-following score for every evaluated model.
    Top scores: Opus 4.8 reaches 88.0, Opus 4.7 87.7, Sonnet 4.6 85.9,
    GLM 5.1 85.0, Gemini 3.5 Flash 81.6, Gemini 3.1 Pro Preview 81.5.
    Bottom scores: Nemotron Nano 30B 25.2, Nemotron Super 120B 46.8.}
  \label{fig:teaser}
\end{teaserfigure}

\maketitle

%% Suppress acmart's running header (which would otherwise still show
%% blank "conference" placeholders) so only the page number remains.
\pagestyle{plain}

%% Override acmart's \flushbottom so columns are typeset at natural
%% height instead of stretching paragraph gaps to align bottoms.
\let\flushbottom\relax
\raggedbottom

%% ----------------------------------------------------------------------
\section{Introduction}

LLM-powered agents are quickly reshaping how people develop software,
examine data, and automate sophisticated workflows. One of the
primary ways to extend an agent's capabilities beyond what the
underlying model learned during training is through \emph{skills}\cite{claudeskills}:
reusable knowledge artifacts that capture domain-specific workflows,
API usage patterns, coding conventions, opinionated workflow choices,
and best practices in a structured form.

Skills are increasingly supported across agent platforms and
registries, enabling users to specialize general-purpose agents for
domains such as scientific research, personal productivity, data
engineering, web development, and infrastructure automation. However,
despite their growing practical importance, it remains unclear how to
rigorously evaluate whether a skill actually improves agent behavior.

This gap is reflected in the current benchmark landscape. Most
existing agent benchmarks measure general task solving, tool use, or
coding ability~\cite{jimenez2024swebench, merrill2026terminalbench, swebenchpro2025, jain2024livecodebench,
yao2024taubench}, but do not focus specifically on how skills change
agent behavior across models. To our knowledge, only a limited number
of works focus on evaluating
skills~\cite{skillsbench2026, liu2026skillsinwild, yang2026skillopt}. While they provide
important initial evidence, they rely on small, fixed sets of
hand-authored tasks, which limits their domain coverage and makes it
difficult to draw conclusions about skill utility across the broad and
heterogeneous space of real-world skills distributed across various
community registries~\cite{skillssh, tesslregistry, skillsmp}.
Crucially, these benchmarks score skills only against their own fixed
task suites; they provide no way to take an arbitrary, newly authored
skill and measure whether it actually improves agent behavior ---
precisely the question a skill author faces in practice.

More importantly, fixed benchmark suites do not answer the practical
question faced by a skill author: \emph{given a newly created skill,
how can one determine whether it improves performance on the tasks it
is intended to support?} This question decomposes into several
related evaluation problems. Does the agent follow the instructions
encoded in the skill? Does the skill provide information, knowledge,
workflows, or preferences that are not already captured by the model?
Does the skill improve task completion, or merely change
surface-level behavior? And can access to the skill allow a smaller
or cheaper model to match the performance of a larger one?

In this work, we propose a scalable framework for evaluating the
utility of agent skills. Given a skill, our method generates realistic
tasks for which the skill should be relevant --- without explicitly
revealing the skill-specific behavior being tested --- together with
custom rubrics tailored to the evaluation goal. These tasks become
evaluation samples that can be adapted to different goals:
(1)~a single agent solves the tasks under controlled conditions, with
and without access to the skill, allowing us to estimate the marginal
value of the skill and to identify cases where it changes agent
behavior in ways that are useful, redundant, or ineffective;
(2)~multiple agents solve the same tasks, allowing us to identify the
right model for a given task; and
(3)~multiple tasks are generated and solved by a fixed agent, helping
to surface weak spots of the skill and the specific behaviors it does
or does not induce --- a capability that, to our knowledge, no prior
skill benchmark provides. We explore all 3 directions in our experiments.

We test our framework on a corpus of 500 real-world open-source skills
sourced from trusted companies and organizations across various
community registries. From this corpus, we generate roughly 1{,}000
realistic evaluation tasks and use them to measure the effect of skills
on instruction-following and goal-completion metrics across multiple
model families and capability levels.

To summarize, our contributions are as follows:
\begin{itemize}
  \item We introduce a scalable framework for evaluating agent skills
    that automatically \emph{synthesizes} realistic, executable
    evaluation tasks.
  \item We conduct a broad empirical study of real-world skills
    across diverse domains, evaluating 19 agent-model configurations
    spanning open-source and proprietary models, and show that skills
    induce measurable changes in agent behavior, particularly in how
    closely models adhere to the workflows and conventions a skill
    encodes.
  \item We show that, beyond aggregate benchmarking, our framework can
    evaluate an \emph{individual} skill in isolation --- a capability
    absent from prior skill benchmarks --- giving skill authors a
    concrete tool to locate weak spots and improve them.
  \item We release to the community a dataset of realistic,
    executable coding tasks generated by our framework.
\end{itemize}

%% ----------------------------------------------------------------------
\section{Evaluation Framework}
\label{sec:framework}

Our framework turns one or more skills, optionally guided by a
user-specified intent, into executable evaluation tasks. Each task is
a realistic user request paired with the required environment, input
artifacts, and hidden rubrics. A series of specialized agents
(environment engineering, task generation, and validation) builds
tasks and checks executability, consistency, and rubric leakage. 

The pipeline can run autonomously end-to-end, but also supports
human-in-the-loop control, allowing users to inspect, modify, or
approve intermediate outputs at each stage allowing to focus on the aspects of the skill that matter most to them. 
In practice, we find that this hybrid mode produces the highest-quality tasks, since human
reviewers can correct ambiguous requirements, refine generated
inputs catch subtle validation failures before the tasks are
finalized and control the difficulty of the tasks. 
In this section we explain the main building blocks of the pipeline and in the Experiment Setup section we discuss the exact steps of the evaluation we conducted.

\begin{figure}[t]
  \centering
  \includegraphics[width=\linewidth]{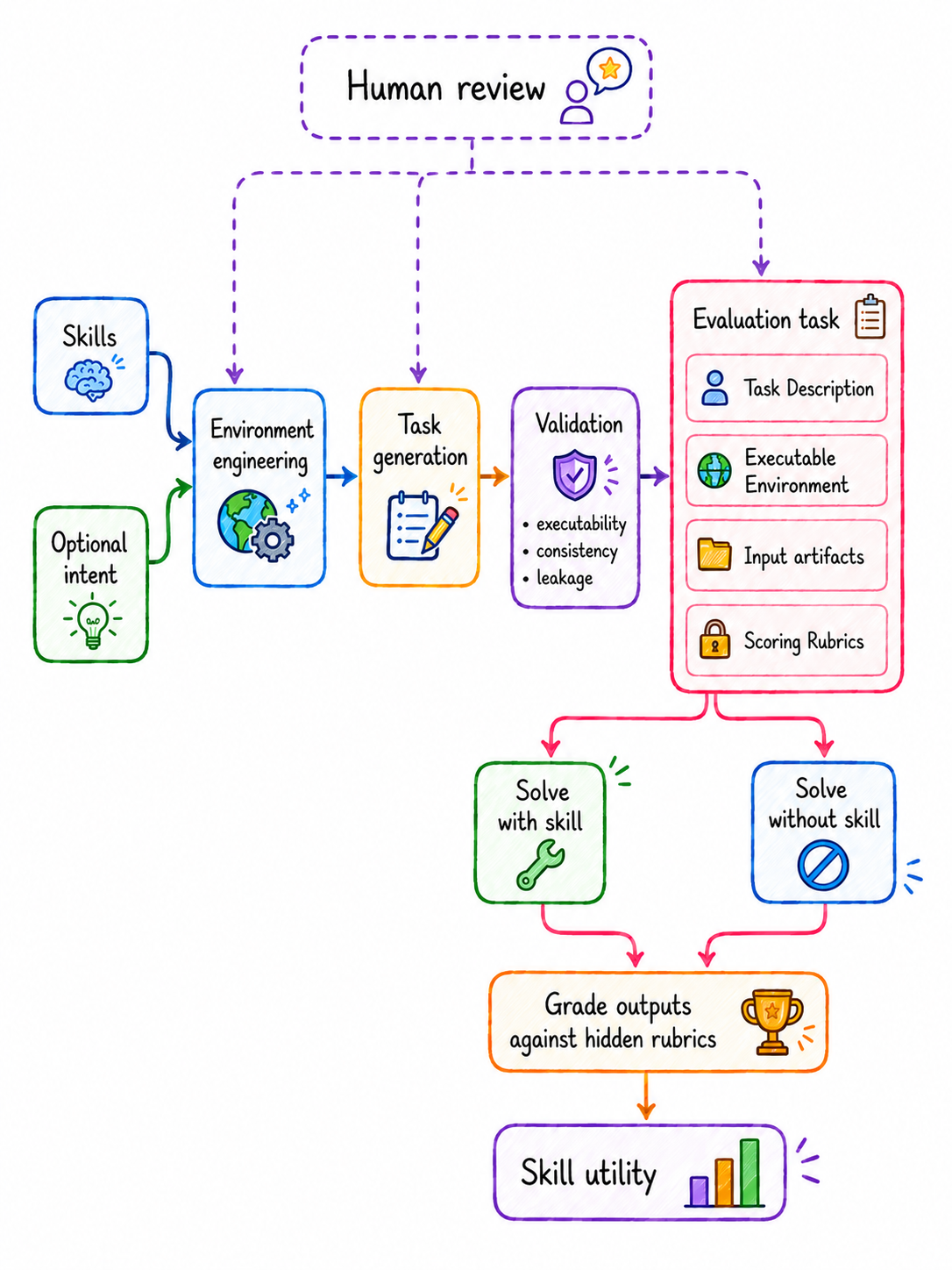}
  \caption{End-to-end overview of the skill evaluation pipeline.
    Skills and an optional user intent are fed to an environment
    engineering agent, then to a task generation agent, and finally
    to a validation agent. Each generated task, if the goal is to measure whether the access to skill substantially changes core model behaviour, can be solved twice ---
    with and without access to the skill --- and the outputs are
    graded against hidden rubrics to produce a per-skill utility
    score. The pipeline supports human review at every stage.}
  \Description{Hand-drawn style diagram of the pipeline: skills and
    optional intent flow into environment engineering, then task
    generation, then validation, then evaluation tasks consisting of
    description, executable environment, input artifacts, and
    scoring rubrics. The tasks are then solved with and without
    skill and outputs are graded.}
  \label{fig:pipeline}
\end{figure}

Figure~\ref{fig:pipeline} summarizes the full pipeline. In the
remainder of this section, we describe each component in detail.

\paragraph{Analyzing skill(s) and user intent.}
To synthesize realistic tasks that reflect real skill usage, we
ideally need a source of user input that clarifies how the skill is
applied in practice, such as a natural-language prompt, Github Issue, a Jira
ticket, or previous user--agent interaction logs that capture the
relevant workflow. However, collecting such inputs at scale is
difficult, and they are typically available only in the
human-in-the-loop setting where users actively engage with the
system. Therefore, in the fully automated setting, we instead rely on
the skill content itself and a quorum of agents that infer realistic
usage scenarios.

\paragraph{Environment engineering.}
Before generating tasks for a given collection of skills, we must
determine whether the execution environment can, in principle, be
constructed. For example, some skills may require access to an
existing database, API tokens, an MCP server~\cite{anthropic2024mcp},
a browser, CLI tools, or a particular framework. Without providing these capabilities at
solve time, the resulting tasks would not be executable. Accordingly,
the goal of the environment engineering agent in our pipeline is to
identify required resources and, in the fully autonomous setting,
attempt to satisfy missing requirements. To constrain the problem, we
categorize environment requirements into the groups listed in
Table~\ref{tab:env-categories}.

\begin{table}[t]
  \caption{Categories of environment dependencies a skill may require.}
  \label{tab:env-categories}
  \small
  \begin{tabular}{p{0.30\linewidth}p{0.60\linewidth}}
    \toprule
    Category & Guidance \\
    \midrule
    Tool and CLI access & Whether required command-line tools are available in the environment. \\
    MCP server access & Whether the skill depends on MCP tooling. \\
    External network access & Whether the skill needs outbound internet / external APIs. \\
    Auth and credential injection & Whether non-public credentials (keys/tokens) are required. \\
    Env variables setup & Whether environment variables must be set for the skill to run. \\
    Runtime / language environment & Whether specific runtimes/system packages are required (e.g., Python/Node/Java). \\
    Multi-turn evaluation support & Whether the skill requires interactive, multi-turn user input. \\
    Existing repository & Whether the skill assumes a repo/project is present to operate on. \\
    Existing code files (outside the skill) & Whether the skill assumes specific files exist outside the skill bundle. \\
    Git repository state & Whether the skill depends on a particular git state (branch/PR/conflict/history). \\
    Access to DB & Whether a database must be reachable for the skill to work. \\
    Browser / UI access & Whether the skill needs a browser or UI automation capability. \\
    Local running services & Whether the skill requires local services to be running (e.g., dev server/queue/cache). \\
    Input file (baseline) & Whether a shared baseline input artifact is required (e.g., PDF/image/data file). \\
    Pre-populated external service state & Whether external services must already contain specific pre-seeded state (tickets/PRs/threads). \\
    Other & Catch-all for dependencies not covered by the listed categories. \\
    \bottomrule
  \end{tabular}
\end{table}

\paragraph{Task generation.}
At this stage of the pipeline, we generate realistic task proposals
derived from the skill content or from user feedback. Some proposals
require additional inputs. For example, a PDF-processing skill
requires different PDFs depending on the scenario. Accordingly, the
task-generation agent also specifies input requirements. In hybrid
mode, a human reviewer can provide missing inputs. In fully
autonomous mode, the agent attempts to obtain them by retrieving
public resources or synthesizing inputs from scratch. The maximum number of
proposals can be set explicitly or determined by the agent depending on the complexity of the skill.

Once inputs are available, the agent converts each proposal into an
executable task specification, including the task description, an
inputs folder, a verifiable execution environment, and scoring
rubrics. Depending on the end goal of the evaluation, the pipeline can generate custom rubrics tailored to the specific aspects of interest, 
this is only possible when external source of feedback is available.
By default, the pipeline produces two rubric sets: one for task
completion and one for instruction following. Each rubric is a
natural-language assertion scored on a 1--10 scale, with scores
summed to 100 per category. We use these rubrics later in our large scale evaluation experiments.

\emph{Task-completion rubrics} assess whether the solution produces
the requested outputs and whether the final artifact is
correct.

\emph{Instruction-following rubrics} assess whether the solution
follows the preferences encoded in the skill, including library
choices, structural conventions, naming rules, prohibited patterns,
and required steps.

\paragraph{Tasks validation.}
Further in the pipeline, we employ a quality-assurance agent to
identify and remove ambiguous tasks and inconsistent environments.
Specifically, we verify that each task includes all required inputs
and that its environment requirements are satisfied. We also check
that the task description does not leak the exact steps needed to
produce the expected solution, and that the task content does not
reveal rubric details that could enable gaming. Based on these
filters, we report quality checks across dimensions such as
environment health and task health. In hybrid mode, a user can
validate and correct specific issues, whereas in autonomous mode we
discard tasks that fail these checks.

\paragraph{Running evaluation.}
At this stage, the pipeline has generated a set of realistic tasks
with fully verifiable environments.
Table~\ref{tab:task-example} shows the structure of one such task,
generated from the Hugging Face \texttt{hf-cli} skill (which migrates
scripts from the legacy \texttt{huggingface-cli} to the current
\texttt{hf} command). Each task bundles a natural-language
description, an inputs folder, and two rubric sets ---
\emph{goal-completion} and \emph{instruction-following} --- that are
hidden from the solver and used only by the judge.

\begin{table*}[t]
  \caption{Structure of a single generated task, illustrated with the
    Hugging Face \texttt{hf-cli} skill. The top section describes the
    artifact handed to the solver; the bottom two sections show
    representative items from the two hidden rubric sets used by the
    judge, with their point budgets.}
  \label{tab:task-example}
  \footnotesize
  \setlength{\tabcolsep}{6pt}
  \renewcommand{\arraystretch}{1.2}
  \begin{tabular}{@{}>{\raggedright\arraybackslash}p{0.18\textwidth}
                     >{\raggedright\arraybackslash}p{0.74\textwidth}
                     r@{}}
    \toprule
    Field & Content & Pts \\
    \midrule
    Skill & \texttt{hf-cli} & \\
    Skill summary &
      Migrate Hugging Face Hub scripts from the deprecated
      \texttt{huggingface-cli} to the current \texttt{hf} command;
      enforces that authentication relies on the \texttt{HF\_TOKEN}
      environment variable rather than a \texttt{login} command or a
      \texttt{-{}-token} flag. & \\
    Task description &
      The script \texttt{inputs/broken\_pipeline.sh} was written with
      an outdated Hugging Face CLI and no longer works. Fix it and
      save to \texttt{solution/fixed\_pipeline.sh}. The fixed script
      must perform the same operations (verify identity, list
      \texttt{my-org} models, download \texttt{my-org/base-model},
      upload \texttt{./datasets/processed} as a dataset, prune
      detached cache revisions, show cache contents), but
      authentication must rely on the \texttt{HF\_TOKEN} environment
      variable. Do not change the workflow --- only the commands. & \\
    Inputs provided &
      \texttt{inputs/broken\_pipeline.sh}: a 7-step bash script that
      uses the deprecated \texttt{huggingface-cli} for every
      operation (\texttt{login -{}-token}, \texttt{whoami},
      \texttt{list models}, \texttt{download}, \texttt{upload},
      \texttt{cache prune}, \texttt{cache list}). & \\
    \midrule
    \multicolumn{3}{@{}l}{\emph{Goal-completion rubric --- 4 of 12 items shown (100 pts total)}} \\
    & \texttt{solution/fixed\_pipeline.sh} is present in the workspace & 10 \\
    & Model listing passes both \texttt{-{}-author my-org} and \texttt{-{}-limit 10} & 10 \\
    & Downloads \texttt{my-org/base-model} with \texttt{-{}-local-dir ./models/base} and \texttt{-{}-revision main} & 10 \\
    & Cache prune runs non-interactively & 5 \\
    \midrule
    \multicolumn{3}{@{}l}{\emph{Instruction-following rubric --- 4 of 16 items shown (100 pts total)}} \\
    & No occurrence of \texttt{huggingface-cli} anywhere in the file & 10 \\
    & Identity verification uses exactly \texttt{hf auth whoami} & 10 \\
    & Authentication relies on \texttt{HF\_TOKEN} only (no \texttt{login} command or \texttt{-{}-token} flag) & 10 \\
    & Every Hub operation uses the \texttt{hf} prefix & 5 \\
    \bottomrule
  \end{tabular}
\end{table*}

A solver agent then attempts to solve each task. In our experiments,
each task is solved twice --- with and without access to the skill.
A separate verification agent scores the resulting solutions against
the rubrics, using an LLM-as-judge setup. During grading, the
verification agent has access to the rubrics, the solver agent's
proposed solutions, and all logs produced by the solver. We analyze
results at the level of individual rubric dimensions. A large
performance delta on goal-completion rubrics suggests that the skill
provides knowledge or capabilities unavailable to the base agent,
preventing it from producing a valid solution for a target task. A
large delta on instruction-following rubrics suggests that the skill
induces behavioral changes in how the agent solves the task. These
two dimensions are then combined into a single weighted score,
reflecting that both matter in practice.

%% ----------------------------------------------------------------------
\section{Dataset Construction}
\label{sec:dataset}

To study the impact of skills, we collected a representative dataset
of high-quality, real-world skills and generated realistic tasks
derived directly from their content. Each task is paired with two
rubric sets: instruction following and goal completion. Our evaluation
aims to measure how different agent--model configurations utilize a
skill, whether access to a skill substantially changes the model's
behavior, and how large the resulting performance gain is. Below, we
describe the dataset construction process. In total, we obtained
approximately 500 unique skills and 1{,}000 tasks derived from these
skills.

\paragraph{Skills collection.}
We source skill metadata from a public skill aggregation
platform,\footnote{\url{https://tessl.io/registry}} focusing on
popular skills from well-known organizations such as Anthropic,
Google, ElevenLabs, and Shopify (approximately 100 organizations in
total). For each skill, we download the full skill folder, including
the \texttt{SKILL.md} file and supporting files, from the original
GitHub repository. We then filter for permissive licenses (MIT and
Apache 2.0) to ensure redistribution rights, remove ill-formatted
skills (e.g., empty names or descriptions), and deduplicate by file
content.

\paragraph{Automatic quality checks.}
In addition to basic filtering, we discard skills that fail Snyk
security checks.\footnote{\url{https://evo.ai.snyk.io/evo-discovery-try-now/}} 
to prevent malicious or harmful skills.

\paragraph{Validating requirements.}
Next, we run the environment engineering stage of our pipeline. The
agent analyzes each skill and extracts its environment requirements,
such as required frameworks, CLI tools, and missing API keys. Table~\ref{tab:env-reqs} shows the resulting
distribution: tool and CLI access, authentication, and a specific
runtime or language environment dominate, while categories such as
database access, browser automation, and pre-populated external
service state each appear in fewer than 10\% of skills. In the
human-in-the-loop setting, missing inputs and environment constraints
can be validated and addressed by a reviewer. In the fully autonomous
setting, some categories are difficult or impossible to satisfy, so
we discard skills that require any of the following: existing
repositories, MCP server access, multi-turn evaluation support,
pre-populated external service state, local running services,
database access, or specific git repository states. We retain skills
that require authentication or credential injection, since many
involve third-party APIs where missing keys do not prevent producing
a valid solution.

\begin{table}[t]
  \caption{Frequency of identified environment requirements across all
    collected skills (a skill may have multiple requirements, so
    columns do not sum to 100\%). The vast majority of skills require
    either a command-line tool, an authentication credential, or a
    specific runtime/language environment.}
  \label{tab:env-reqs}
  \small
  \setlength{\tabcolsep}{6pt}
  \renewcommand{\arraystretch}{1.15}
  \begin{tabular}{@{}>{\raggedright\arraybackslash}p{0.70\linewidth} r@{}}
    \toprule
    Category & Frequency \\
    \midrule
    Tool and CLI access            & 70.4\% \\
    Auth and credential injection  & 66.0\% \\
    Runtime / language environment & 65.2\% \\
    External network access        & 42.0\% \\
    Existing repository            & 19.3\% \\
    MCP server access              & 16.2\% \\
    Multi-turn evaluation support  &  7.0\% \\
    Other (pre-populated external service state,
      env-variables setup, existing code files outside the skill,
      browser/UI access, local running services, access to DB,
      input file, git repository state, other miscellaneous) & ${\le}\,4.4\%$ each \\
    \bottomrule
  \end{tabular}
\end{table}

\begin{figure}[t]
  \centering
  \includegraphics[width=\linewidth]{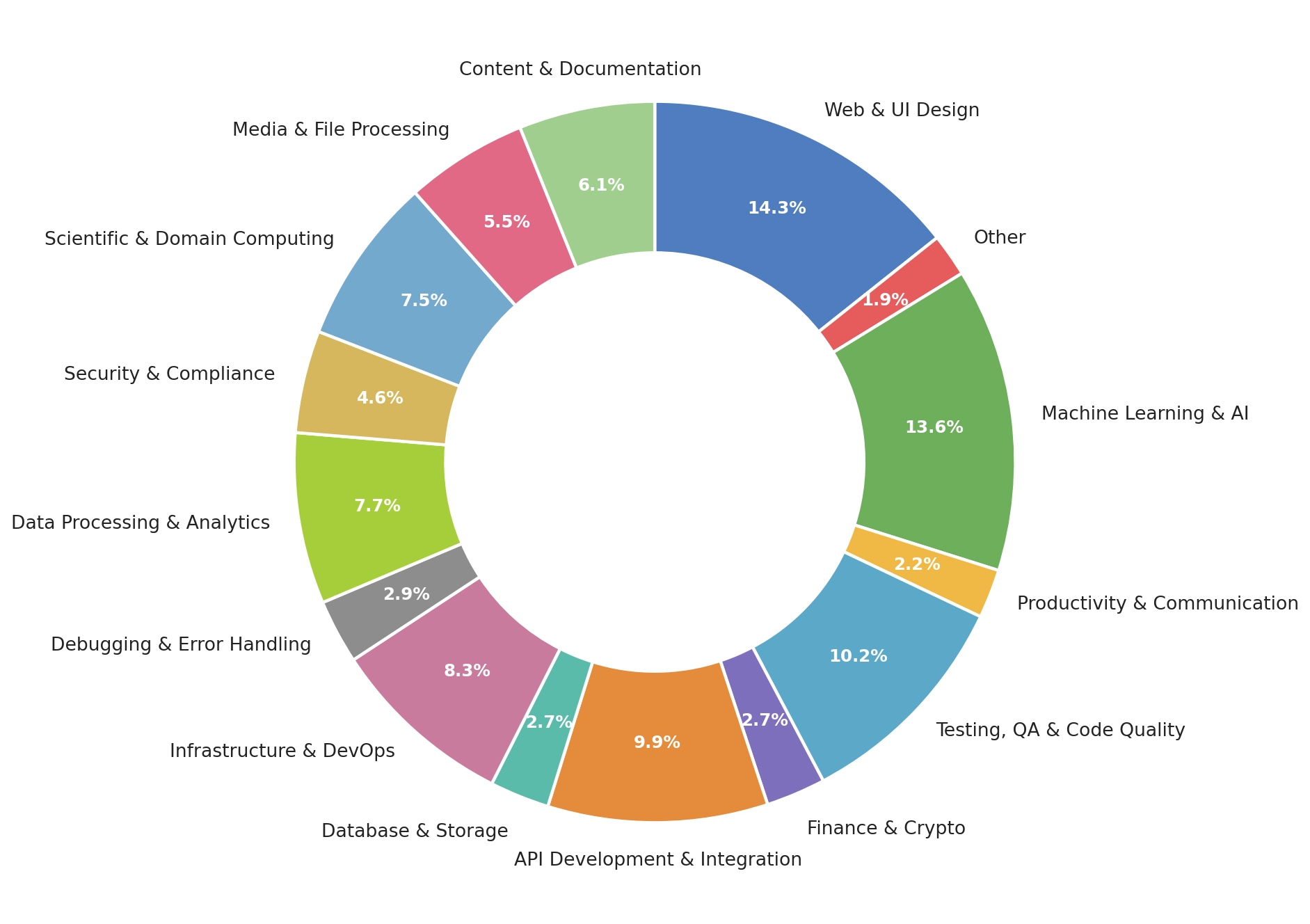}
  \caption{Distribution of skills across high-level themes obtained
    by clustering.}
  \Description{Donut chart of skill clusters: Web \& UI Design
    14.3\%, Machine Learning \& AI 13.6\%, Testing, QA \& Code
    Quality 10.2\%, Finance \& Crypto 9.9\%, Infrastructure \&
    DevOps 8.3\%, Data Processing \& Analytics 7.7\%, Scientific \&
    Domain Computing 7.5\%, Content \& Documentation 6.1\%, Media \&
    File Processing 5.5\%, Security \& Compliance 4.6\%, Database
    \& Storage 2.9\%, API Development \& Integration 2.7\%,
    Debugging \& Error Handling 2.7\%, Productivity \&
    Communication 2.7\%, Other 1.9\%.}
  \label{fig:clusters}
\end{figure}

\paragraph{Clustering analysis.}
After filtering, we obtain a dataset of approximately 500 skills.
We perform clustering to identify high-level themes. Most clusters
relate to programming and software engineering, including web and UI
design, machine learning and AI, infrastructure and DevOps, testing
and code quality, and API development. The remaining clusters are
broadly associated with science and research, finance and marketing,
content and documentation, and personal productivity.
Figure~\ref{fig:clusters} shows the breakdown.

\paragraph{Task generation.}
Finally, we run the task-generation stage in fully autonomous mode,
in which the agent resolves missing inputs when possible. Each
validated task includes a task description, an optional inputs
folder, and scoring rubrics. We generate up to three diverse tasks
from each skill or combination of skills. A quality-assurance agent
then validates that tasks are executable and that the task description
does not leak rubric content; tasks that fail these checks are
discarded, so fewer may remain per skill. Because skills from trusted providers are often
organized around natural themes, we generate both single-skill tasks
and multi-skill tasks. This procedure yields approximately 1{,}000
tasks.

%% ----------------------------------------------------------------------
\section{Experimental Setup}
\label{sec:setup}

We evaluated several agent harnesses, including both commercial and
open-source options, across 19 frontier models, both open-source and
closed-source, under two skill conditions. This resulted in
approximately 38{,}000 valid trajectories. A trajectory is valid when
the agent passes, fails, or times out on a task without infrastructure
or runtime errors. Each trajectory is one agent's attempt at solving
a single task under a specific skill condition. We scored each valid
trajectory with the LLM-as-judge approach against two concrete scoring
rubrics.

\paragraph{Agent harnesses.}
We evaluate two closed-source agent harnesses, Claude Code~\cite{claudecode}
and Codex CLI~\cite{openaicodex}, on their corresponding model families.
For the remaining models we use the open-source OpenHands
harness~\cite{wang2025openhands}, as well as Claude Code itself, which
supports custom model backends, on a subset of closed- and open-source
models.

\paragraph{Models.}
We evaluate commercial models from Anthropic (Haiku 4.5~\cite{haiku45},
Sonnet 4.6~\cite{sonnet46}, Opus 4.7~\cite{opus47}, Opus
4.8~\cite{opus48}), OpenAI (GPT-5.4 nano~\cite{gpt54nano}, GPT-5.4
mini~\cite{gpt54mini}, GPT-5.4~\cite{gpt54}), and Google (Gemini 3 Flash
Preview~\cite{gemini3flashpreview}, Gemini 3.1 Flash
Lite~\cite{gemini31flashlite}, Gemini 3.1 Pro
Preview~\cite{gemini31propreview}, Gemini 3.5
Flash~\cite{gemini35flash}) serving directly from the model providers. We also evaluate frontier open-source
models, including GLM 5.1~\cite{glm51}, DeepSeek V4
Pro and Flash~\cite{deepseekv4pro}, Kimi K2.6~\cite{kimik26}, MiniMax
2.7~\cite{minimax27}, Qwen3-Coder-Next~\cite{qwen3codernext}, and the
NVIDIA Nemotron 3 series (Super 120B~\cite{nemotron3super120b} and Nano
30B~\cite{nemotron3nano30b}), served via Fireworks
AI\footnote{\url{https://fireworks.ai/models}} or Amazon
Bedrock\footnote{\url{https://aws.amazon.com/bedrock/}}.

\paragraph{Skills conditions.}
We evaluate each task under two conditions.
\begin{itemize}
  \item \emph{Without skill.} The agent receives the task description
    and a valid execution environment. The skill is not installed in the execution environment.
  \item \emph{With skill.} The agent receives the task description
    and a valid execution environment. In addition, the agent is
    explicitly informed that the relevant skills are installed and
    available.
\end{itemize}
Such \emph{With skill} design isolates skill utility by reducing cases where a skill is available but unused because the agent fails to recognize its relevance. 
In realistic deployment settings, users do not necessarily indicate which skill should be used; 
here, however, this assumption allows us to decouple the utility of a skill once invoked from the agent’s ability to recognize its relevance.

\paragraph{Metrics.}
During task generation, we produce two rubric sets: task-completion
rubrics and instruction-following rubrics. Each set sums to 100
points. We report each metric individually as well as a weighted
average of the two --- which metric matters more depends on the
specific use case. We also report the skill delta improvement (the
difference between the with-skill and without-skill scores), runtime,
cost, and token consumption for each model.

\paragraph{Evaluation protocol.}
We use Sonnet 4.6 in the Claude Code harness as the judge agent
across all experiments. The judge scores each solution against both
rubric sets, producing a separate score for each. Logs from the
solver agent are also provided to the judge.

%% ----------------------------------------------------------------------
\section{Results}
\label{sec:results}

Our results are twofold. First, we evaluate the impact of skills on
realistic tasks across multiple frontier model families, examining
how closely models adhere to the skill and how much their performance
improves once they have access to it. Second, we analyze performance
across different skill categories and inspect specific skills in
detail. Table~\ref{tab:per-model} provides a fine-grained breakdown of
performance for all models across rubrics, runtime, cost, and token
consumption.

\subsection{Relative Performance Gains Across Model Families}
Across every model we evaluate, access to a relevant skill yields a
substantial improvement in the instruction-following and overall
scores, with relative gains ranging from 5.5 to 22 points depending
on the model and driven primarily by the instruction-following
component. Such an improvement is expected by construction: the delta
compares the with-skill and without-skill conditions, which inherently
favors the with-skill setup. Nevertheless, this result demonstrates
that skills consistently express opinionated choices that shift the
model's behavior relative to its base knowledge.

However, models benefit from a skill to very different degrees. Models
from the Nemotron family barely benefit at all, and, more surprisingly,
the recent Kimi K2.6 also fails to capitalize on access to the skill,
yielding only a 7.1-point boost. This indicates that the model does
not utilize the skill's content properly, relying on its own
capabilities rather than the provided context.

Overall, we observe that the relative impact of skills is larger for
smaller models than for bigger ones --- visible in Haiku and Sonnet
versus the Opus models, GPT-5.4 nano versus GPT-5.4, and Gemini 3
Flash Preview versus Gemini 3.1 Pro Preview.

\begin{table*}[!t]
  \caption{Per-model breakdown of instruction following, goal
    completion, and overall score (a weighted average of the two),
    along with the skill delta ($\Delta$, the gain in overall score
    from adding the skill), runtime, cost, and token consumption, in
    the with-skill (w/) and without-skill (w/o) conditions. Anthropic
    models run in the Claude Code harness, OpenAI models in Codex, and
    the remaining models in Claude Code (DeepSeek, Kimi, GLM, MiniMax)
    or OpenHands (Gemini, Nemotron, Qwen). Tokens are rounded to the
    nearest thousand.}
  \label{tab:per-model}
  \footnotesize
  \setlength{\tabcolsep}{3pt}
  \renewcommand{\arraystretch}{1.1}
  \begin{tabular}{@{}l rr rr rr r rr rr rr rr@{}}
    \toprule
    & \multicolumn{2}{c}{Instruction Following}
    & \multicolumn{2}{c}{Goal Completion}
    & \multicolumn{2}{c}{Overall Score}
    & {Skill $\Delta$}
    & \multicolumn{2}{c}{Runtime (min)}
    & \multicolumn{2}{c}{Cost (\$/scenario)}
    & \multicolumn{2}{c}{Input Tokens (k)}
    & \multicolumn{2}{c}{Output Tokens (k)} \\
    \cmidrule(lr){2-3} \cmidrule(lr){4-5} \cmidrule(lr){6-7}
    \cmidrule(lr){8-8} \cmidrule(lr){9-10} \cmidrule(lr){11-12}
    \cmidrule(lr){13-14} \cmidrule(lr){15-16}
    Model & w/o & w/ & w/o & w/ & w/o & w/ & & w/o & w/ & w/o & w/ & w/o & w/ & w/o & w/ \\
    \midrule
    Opus 4.8                & 59.8 & 88.0 & 93.3 & 97.5 & 76.6 & 92.7 & +16.2 & 2.7 & 2.4 & 2.66 & 3.26 &  471 &  595 & 12.3 & 11.5 \\
    Opus 4.7                & 56.8 & 87.7 & 91.7 & 96.9 & 74.2 & 92.3 & +18.1 & 2.1 & 2.3 & 2.56 & 3.94 &  470 &  743 &  8.4 &  8.9 \\
    Sonnet 4.6              & 49.0 & 85.9 & 89.8 & 97.0 & 69.4 & 91.5 & +22.1 & 2.2 & 2.5 & 1.07 & 1.46 &  318 &  447 &  7.8 &  8.2 \\
    Haiku 4.5               & 43.6 & 75.3 & 85.3 & 93.0 & 64.4 & 84.1 & +19.7 & 1.0 & 1.3 & 0.08 & 0.11 &  347 &  464 &  5.8 &  6.6 \\
    GPT-5.4                 & 56.5 & 79.9 & 92.0 & 96.5 & 74.2 & 88.2 & +13.9 & 3.9 & 3.8 & 1.04 & 1.10 &  664 &  617 & 14.9 & 14.9 \\
    GPT-5.4 mini            & 49.8 & 74.5 & 89.5 & 94.5 & 69.7 & 84.5 & +14.8 & 1.7 & 1.7 & 0.71 & 0.81 &  411 &  456 & 11.5 & 11.6 \\
    GPT-5.4 nano            & 40.8 & 70.4 & 85.3 & 93.4 & 63.0 & 81.9 & +18.9 & 3.0 & 3.2 & 0.07 & 0.08 &  737 &  802 & 16.1 & 17.2 \\
    Gemini 3.5 Flash        & 55.1 & 81.6 & 91.1 & 95.6 & 73.1 & 88.6 & +15.5 & 4.0 & 3.7 & 1.17 & 1.19 & 1540 & 1610 & 25.7 & 24.1 \\
    Gemini 3.1 Pro Preview  & 51.9 & 81.5 & 89.0 & 95.6 & 70.4 & 88.5 & +18.1 & 4.2 & 3.5 & 0.87 & 0.70 &  863 &  688 & 12.3 &  9.6 \\
    Gemini 3 Flash Preview  & 46.1 & 78.2 & 85.7 & 93.6 & 65.9 & 85.9 & +20.0 & 3.2 & 3.8 & 0.11 & 0.14 &  429 &  668 & 11.8 & 11.2 \\
    Gemini 3.1 Flash Lite   & 37.3 & 58.1 & 79.3 & 84.7 & 58.3 & 71.4 & +13.0 & 0.8 & 1.1 & 0.03 & 0.04 &  210 &  314 &  2.6 &  3.1 \\
    DeepSeek V4 Pro         & 48.0 & 77.5 & 88.3 & 95.6 & 68.1 & 86.6 & +18.4 & 3.7 & 3.9 & 0.65 & 0.75 &  477 &  594 &  8.1 &  8.1 \\
    DeepSeek V4 Flash       & 46.6 & 73.2 & 88.3 & 94.7 & 67.4 & 83.9 & +16.5 & 3.7 & 4.0 & 0.64 & 0.76 &  498 &  619 &  8.9 &  9.2 \\
    Kimi K2.6               & 48.7 & 60.1 & 89.3 & 92.0 & 69.0 & 76.1 &  +7.1 & 2.9 & 2.8 & 0.74 & 0.73 &  605 &  592 & 12.6 & 11.9 \\
    GLM 5.1                 & 51.2 & 85.0 & 90.3 & 97.2 & 70.7 & 91.1 & +20.3 & 4.2 & 3.4 & 0.72 & 0.89 &  487 &  611 &  8.9 &  7.3 \\
    Nemotron 3 Super 120B   & 30.4 & 46.8 & 66.0 & 66.0 & 48.2 & 56.4 &  +8.2 & 4.3 & 4.4 & 0.07 & 0.08 &  436 &  516 &  5.3 &  5.2 \\
    Nemotron 3 Nano 30B     & 18.6 & 25.2 & 45.7 & 50.0 & 32.1 & 37.6 &  +5.5 & 2.7 & 3.2 & 0.03 & 0.04 &  520 &  639 &  5.7 &  6.6 \\
    MiniMax 2.7             & 40.1 & 59.0 & 84.8 & 89.2 & 62.4 & 74.1 & +11.7 & 2.9 & 3.3 & 0.46 & 0.53 &  398 &  496 &  5.7 &  5.5 \\
    Qwen3-Coder-Next        & 36.3 & 57.3 & 78.5 & 84.3 & 57.4 & 70.8 & +13.4 & 7.9 & 8.2 & 0.47 & 0.50 &  919 &  979 &  9.9 &  9.2 \\
    \bottomrule
  \end{tabular}
\end{table*}

\subsection{Behavioral Effects: Instruction Following Versus Goal Completion}

As table~\ref{tab:per-model} shows, in terms of goal completion, almost all models solve the tasks in both
conditions, with and without skills, and access to a skill pushes goal
completion close to saturation, frequently exceeding 90\%. The main
exception is the Nemotron family, which performs poorly; we attribute
this to its smaller model size or to a lack of relevant data in its
training recipe. Manual inspection suggests two broader explanations
for why goal completion saturates: (i)~our data-synthesis method is
not perfectly calibrated to discriminate the goal-completion
capabilities of frontier models and would require more compute during
task generation to do so; and (ii)~frontier models have become
powerful enough that constructing a truly challenging task in a fully
synthetic setup is difficult, and only externally provided feedback or
human-in-the-loop collaboration can address this.

The instruction-following rubric behaves differently, leading to
substantial variance in overall scores across models.
Figure \ref{fig:teaser} summarizes the results across models. For example, Kimi K2.6, Qwen3-Coder-Next,
MiniMax 2.7, and the older Gemini 3.1 Flash Lite show a visible gap
--- roughly 20--30 points lower in instruction following --- relative
to the strongest models in the Anthropic, OpenAI, and DeepSeek
families, the larger Gemini variants, and GLM 5.1; the Nemotron family
lags by an even wider margin. This gap may indicate a lack of relevant
data in these models' training recipes.

A model that formally achieves the goal but follows instructions poorly can be problematic in real-world applications: ignoring an
instruction may degrade the quality, efficiency, safety, and
reusability of the produced code --- properties that matter for
downstream maintenance even when the immediate task is solved.

\subsection{Skills Close the Gap Between Cheaper and Frontier Models}

A consistent pattern in Table~\ref{tab:per-model} is that, once every
model has access to a relevant skill, the smaller members of a family
become competitive with its largest one. With a skill, GPT-5.4 mini
reaches an overall score of 84.5, only a few points behind the full
GPT-5.4 at 88.2; the same holds for DeepSeek, where V4 Flash (83.9)
trails V4 Pro (86.6) by under three points. The skill narrows the
within-family gap to the point where the cheaper model becomes a
competitive substitute for its flagship.

The effect carries across the open-source/commercial divide. GLM 5.1,
an open-weights model, reaches 91.1 with a skill --- essentially
matching Sonnet 4.6 (91.5) and trailing the top-scoring proprietary
models, Opus 4.8 (92.7) and Opus 4.7 (92.3), by only 1.2--1.6 points.
Other open models also land within a few points of the frontier once a
relevant skill is available (DeepSeek V4 Pro at 86.6, Gemini 3.5 Flash
at 88.6). Because skills lift cheaper models to near-frontier quality,
reaching a given score does not require the most expensive model:
GLM 5.1 attains 91.1 at roughly \$0.89 per scenario, against \$1.46
for Sonnet 4.6 at 91.5, \$3.26 for Opus 4.8 at 92.7, and \$3.94 for
Opus 4.7 at 92.3 --- comparable quality at three- to four-fold lower
cost than the Opus tier. For workloads where the relevant skills are
known in advance --- or where specific tasks can be delegated to
cheaper models --- pairing a skill with a cheaper model is an
attractive alternative to overpaying for a larger commercial system.

\subsection{What Types of Skills Are Most Impactful}

\begin{table}[t]
  \caption{Mean instruction-following score (\%) by skill domain,
    aggregated across all model--harness combinations, in the
    without-skill (w/o) and with-skill (w/) conditions. The rightmost
    column shows the percentage-point uplift contributed by the skill.
    Domains are sorted by uplift, largest first.}
  \label{tab:per-domain}
  \small
  \setlength{\tabcolsep}{6pt}
  \renewcommand{\arraystretch}{1.15}
  \begin{tabular}{@{}>{\raggedright\arraybackslash}p{0.50\linewidth} r r r@{}}
    \toprule
    Domain & w/o Skills & w/ Skills & Uplift \\
    \midrule
    Media \& File Processing        & 32.2 & 70.3 & \textbf{+38.1} \\
    Productivity \& Communication   & 20.0 & 52.5 & \textbf{+32.5} \\
    Security \& Compliance          & 48.1 & 78.4 & \textbf{+30.3} \\
    Content \& Documentation        & 35.2 & 65.5 & \textbf{+30.3} \\
    Database \& Storage             & 41.6 & 69.5 & \textbf{+27.9} \\
    Machine Learning \& AI          & 37.0 & 64.1 & \textbf{+27.2} \\
    Debugging \& Error Handling     & 51.6 & 77.8 & \textbf{+26.2} \\
    API Development \& Integration  & 49.4 & 75.2 & \textbf{+25.9} \\
    Web \& UI Design                & 49.4 & 73.9 & \textbf{+24.6} \\
    Infrastructure \& DevOps        & 55.4 & 77.5 & \textbf{+22.1} \\
    Finance \& Crypto               & 45.2 & 67.0 & \textbf{+21.9} \\
    Data Processing \& Analytics    & 27.6 & 46.6 & \textbf{+19.0} \\
    Scientific \& Domain Computing  & 47.2 & 64.2 & \textbf{+17.0} \\
    Testing, QA \& Code Quality     & 52.2 & 68.9 & \textbf{+16.7} \\
    \bottomrule
  \end{tabular}
\end{table}

We observe that the largest gains accrue in domains where skills encode
specific workflows --- that is, where the skill spells out \emph{how} a
task should be performed; Table~\ref{tab:per-domain} summarizes these
findings. Media \& File Processing sees the biggest
uplift (+38.1) because it is full of skills that involve media-file
editing (e.g., video trimming, audio conversion, image processing,
audiobook generation), all of which require strict adherence to a
particular format, sequence, or convention. The same pattern holds in
Security \& Compliance (+30.3), where skills package up checklists,
exact CLI invocations, and report schemas --- clear steps that are easy
to follow but hard to invent from scratch. The smaller gains appear in
categories that mainly declare guidelines, best practices, and
recommendations --- reasoning-heavy rather than procedural. This
includes Testing, QA \& Code Quality and Data Processing \& Analytics,
whose skills describe general principles rather than concrete
procedures. This suggests a simple heuristic: when knowledge can be
captured as a workflow, it is a strong candidate for a skill.

\subsection{Diagnosing and Improving Skills}
\label{sec:individual}

Our evaluation framework can be used not just to produce an
aggregate score, but also to drill down into specific skills. On one
hand, by generating multiple tasks, different aspects of a skill can be
assessed. On the other hand, inspecting the rubric criteria where the
gap between the with-skill and without-skill solutions is largest lets
us identify the concrete behaviors a skill induces or prevents.
Together, these approaches make it possible to determine, for any
individual skill or combination of skills, whether it changes agent
behavior and, if so, in which way.

We illustrate this last point --- inspecting specific rubrics --- on a
representative skill from the Hugging Face
collection: the \texttt{hf-cli} skill, which teaches the agent to use
the new CLI name \texttt{hf} rather than the legacy
\texttt{huggingface-cli} prefix. In the without-skill condition the
agent reliably falls back to \texttt{huggingface-cli}. The
skill flips this behavior across multiple sub-commands.
Table~\ref{tab:hf-cli} summarizes the key rubric criteria where the
two solutions diverge.

\begin{table*}[t]
  \caption{Selected rubric criteria for the \texttt{hf-cli} Hugging
    Face skill, contrasting the with-skill and without-skill
    solutions for a single representative task.}
  \label{tab:hf-cli}
  \small
  \renewcommand{\arraystretch}{1.15}
  \begin{tabular}{@{}>{\raggedright\arraybackslash}p{0.40\textwidth}
                     >{\raggedright\arraybackslash}p{0.20\textwidth}
                     >{\raggedright\arraybackslash}p{0.27\textwidth}
                     r@{}}
    \toprule
    Rubric requirement & With skill & Without skill & Gain \\
    \midrule
    Solution contains no occurrence of \texttt{huggingface-cli} anywhere
      & 0 occurrences
      & Uses \texttt{huggingface-cli} as the prefix for every Hub operation
      & +10 \\
    Identity verification uses \texttt{hf auth whoami}
      & \texttt{hf auth whoami} present
      & Uses \texttt{huggingface-cli whoami}
      & +10 \\
    Push to Hub uses \texttt{hf upload}
      & \texttt{hf upload} present
      & Uses \texttt{huggingface-cli upload}
      & +5 \\
    Model listing uses \texttt{hf models list} (not \texttt{hf list models})
      & \texttt{hf models list} present
      & Uses \texttt{huggingface-cli list}
      & +5 \\
    \bottomrule
  \end{tabular}
\end{table*}

Our evaluation framework makes it quick and easy to identify where
agent behavior is changed due to access to a skill and where it is
not, and, more importantly, provides a mechanism for measuring the
value of each individual component of the skill and whether it's adherent to the skill's content.

%% ----------------------------------------------------------------------
\section{Discussion}
\label{sec:discussion}

\paragraph{What does ``skill utility'' actually measure?}
Modern frontier models have become so powerful that the formal ability
to complete a task matters less than \emph{how} the task is solved. The
opinionated instructions that capture this \emph{how} can be encoded in
a skill, which lets us rigorously build new evaluations in which the
relevant question in 2026 is not ``can the agent solve this task?'' but
``does the agent solve the task the way I want it to?''. The
large-scale evaluations we conduct in this work are a first step in
this direction.

\paragraph{Implications for skill authors.}
For skill authors, the ability to generate realistic tasks with a clean
evaluation methodology and to specify custom rubrics
(Section~\ref{sec:individual}) is arguably more actionable than the
aggregate score. Even the default instruction-following and
goal-completion rubrics used in this study are useful in practice.
When an agent solves a task with and without the skill yet shows a
negligible delta, it likely already captures the required behavior, so
the skill can be removed; conversely, a large delta marks the parts of
the skill that are actually doing the work.

\paragraph{Implications for deployment.}
The fact that cheap models with skills routinely match expensive
models without them has direct deployment consequences. For
production workloads where the relevant skills are known in advance,
serving a smaller model and prepending the skill content is often a
better operating point than serving a larger model without any
guidance, both in dollars and in tokens. Our cost numbers in
Table~\ref{tab:per-model} make this concrete.

\paragraph{Limitations.}
Several limitations are worth noting. First, our with-skill condition
discloses skill relevance to the agent. In real usage, skills might
be installed but never selected. We expect that the gap between the
two settings will be a function of the agent's skill-selection
ability, which deserves its own study. Second, our evaluation depends
on an LLM-as-judge with a single judge model (Sonnet 4.6). While the
rubrics are concrete and the judge generally agrees with manual spot
checks, judge bias remains a concern, especially on aesthetic or
stylistic criteria. Third, the dataset is biased toward
software-engineering domains because that is where most publicly
released skills currently live; we expect the framework to be useful
in other domains, but our empirical claims should be read with that
distribution in mind. Finally, we filter out skills that require
hard-to-reproduce environments (databases, MCP servers, multi-turn
interaction, pre-populated state), which excludes a non-trivial
slice of real-world skills; extending the framework to those
categories is an important direction for future work.

%% ----------------------------------------------------------------------
\section{Related Work}
\label{sec:related}

Our work connects to three lines of prior research: (1) benchmarking
LLM-based agents, (2) augmenting agents with procedural knowledge and
tools, and (3) methodologies for measuring and comparing skill-driven
improvements across a wide range of models.

\paragraph{Reusable knowledge for LLM agents.}
A growing body of work extends LLM agents beyond what the base model
learned in pre-training by supplying additional context, structure, or
capabilities at inference time. Earlier approaches shape the agent's
reasoning process --- chain-of-thought prompting~\cite{wei2022cot} and
ReAct~\cite{yao2023react} interleave reasoning and actions for
multi-step problem solving --- or attach external knowledge and tools,
as in retrieval-augmented generation~\cite{lewis2020rag} and tool-use
interfaces~\cite{yao2024taubench}. A related line of work studies how
agents can accumulate \emph{reusable} knowledge across tasks:
Voyager~\cite{wang2023voyager}
grows a skill library through exploration in an embodied environment,
cognitive-architecture frameworks such as CoALA~\cite{sumers2023coala}
organize memory, actions, and decision-making into reusable
components, and other work encodes API-usage knowledge for popular
libraries into structured Markdown files that help agents navigate
large codebases~\cite{abstraction-adherence}. \emph{Skills} are the
most recent and increasingly standardized incarnation of this idea,
packaging domain-specific workflows, conventions, and API usage into
portable artifacts that can be attached to any
agent~\cite{claudeskills, anthropic2025agentskills}. These efforts
focus largely on \emph{creating} or structuring reusable knowledge;
our work instead addresses the complementary question of whether a
given skill actually changes agent behavior, and provides a scalable
method to measure this for any individual skill.

\paragraph{Skill-aware benchmarks.}
To our knowledge, \textsc{SkillsBench}~\cite{skillsbench2026} is the only
public benchmark designed specifically to test whether a skill
changes agent behavior. It pairs a small number of skills with
hand-authored tasks and grades solutions with unit tests. We adopt a
similar with-skill/without-skill contrast,
but generate tasks \emph{programmatically} and grading rubrics from skills, which lets us
scale from $\sim$90 to thousands of tasks while still controlling for
leakage and quality through automated validation. 

\paragraph{Automated task generation.}
Several recent works generate environments and tasks automatically from
documentation, code, or model rollouts. MCP-Bench~\cite{wang2025mcpbench}
uses a similar synthetic-data-generation approach to construct realistic,
solvable tasks for real-world MCP servers, and scores solutions with an
LLM-as-judge protocol akin to ours. Recent work on GitHub-issue
resolution extends this idea by relying on autonomous quality-assurance
agents to verify that each constructed environment is executable and
healthy\cite{badertdinov2025swerebench, badertdinov2026swerebenchv2, qwen3codernext}.

\paragraph{LLM-as-judge.}
We use the LLM-as-judge protocol~\cite{zheng2023llmjudge}, extending it
to an agentic setting with a fixed, strong judge, and build on broader
guidance for designing AI agent
evaluations~\cite{anthropic2026demystifyingevals, bloom2025}. Recent
work on agentic rubrics~\cite{raghavendra2026agenticrubrics} shows
that rubrics derived from existing context --- a codebase, or, in our
case, a skill --- yield scores that are consistent with ground-truth
tests while also flagging issues that the tests themselves do not
capture.

\section{Conclusion}
\label{sec:conclusion}

We presented a scalable framework for evaluating the utility of agent
skills. In our experiments, given a skill, the framework \emph{synthesizes} realistic,
executable tasks from the skill's content, builds
verifiable environments, and grades solutions against hidden rubrics.
Applied to a corpus of approximately 500 real-world open-source skills
and 1{,}000 generated tasks, we evaluated 19 agent--model
configurations along both instruction-following and goal-completion
axes, and observed several consistent patterns:
(i) access to a relevant skill yields aggregate improvements of
5--22 points, driven largely by instruction following;
(ii) models vary widely in how closely they adhere to the
instructions encoded in a skill;
(iii) once a skill is available, the gap between smaller and
flagship models within a family narrows substantially, making
skill-augmented inference with cheaper models an attractive
deployment option; and (iv) the largest gains accrue in domains
whose skills encode concrete workflows rather than general
best-practice guidelines. 

Beyond aggregate scores, the framework can
evaluate an \emph{individual} skill in isolation --- a capability
absent from prior skill benchmarks --- giving skill authors a
concrete tool for creating custom evaluations, diagnosing weaknesses,
and improving the behaviors they intend to teach. We release our
evaluation dataset to support future work on agent skills.

\bibliographystyle{ACM-Reference-Format}
\bibliography{references}

\end{document}